\documentclass[12pt]{article}
\usepackage{graphicx}
\usepackage{graphics}
\title{Double series expression for the Stieltjes constants} 
\author{Mark W. Coffey\\
Department of Physics\\
Colorado School of Mines\\
Golden, CO  80401\\
(Received $\mbox{~~~~~~~~~~~~~~~~~~~~~~~~~~~~~~~2011}$)}
\date{January 25, 2011}
\begin{document}
\maketitle
\baselineskip=25 pt
\begin{abstract}

We present expressions in terms of a double infinite series for the Stieltjes
constants $\gamma_k(a)$.  These constants appear in the regular part of the
Laurent expansion for the Hurwitz zeta function.  We show that the case
$\gamma_k(1)=\gamma$ corresponds to a series representation for the Riemann zeta function given much earlier by Brun.  As a byproduct, we obtain a parameterized
double series representation of the Hurwitz zeta function.

\end{abstract}
 
\medskip
\baselineskip=15pt
\centerline{\bf Key words and phrases}
\medskip 

\noindent

Hurwitz zeta function, Stieltjes constants, series representation, Riemann zeta function, Laurent expansion, digamma function 

\vfill
\centerline{\bf 2010 AMS codes} 
11M06, 11Y60, 11M35

\baselineskip=25pt
\pagebreak
\medskip
\centerline{\bf Introduction and statement of results}
\medskip

Herein we supplement our very recent work \cite{coffeyjnt} showing how a method of Addison \cite{addison} for series representations of the Euler constant $\gamma$ may be generalized in many different ways.  We focus on the Stieltjes constants $\gamma_k(a)$
and recall the defining Laurent expansion \cite{coffeyjmaa,coffeystdiffs,coffey2009,stieltjes,wilton2}
$$\zeta(s,a)={1 \over {s-1}}+\sum_{n=0}^\infty {{(-1)^n} \over {n!}}\gamma_n(a)(s-1)^n,
~~~~~~s \neq 1.  \eqno(1.1)$$
where $\zeta(s,a)$ is the Hurwitz zeta function.  For Re $s>1$ and Re $a>0$ we have
$$\zeta(s,a)=\sum_{n=0}^\infty {1 \over {(n+a)^s}}, \eqno(1.2)$$
and by analytic continuation $\zeta(s,a)$ extends to a meromorphic function through out
the whole complex plane.  Notationally, we let $\zeta(s)=\zeta(s,1)$ be the Riemann zeta function \cite{edwards,ivic,riemann,titch} and $\psi=\Gamma'/\Gamma$ be the digamma function (e.g., \cite{nbs,andrews,grad}) with the Euler constant $\gamma=\gamma_0(1)=-\psi(1)$. 
In addition, $P_1(x)=B_1(x-[x])=x-[x]-1/2$ denotes the first periodized Bernoulli polynomial, with $\{x\}=x-[x]$ the fractional part of $x$.  

The sequence $\{\gamma_k(a)\}_{k=0}^\infty$ exhibits complicated changes in
sign with $k$.  For instance, for both even and odd index, there are infinitely
many positive and negative values.  Furthermore, there is sign variation with the
parameter $a$.  These features, as well as the exponential growth in magnitude
in $k$, are now well captured in an asymptotic expression (\cite{knessl2}, Section 2).
In fact, though initially derived for large values of $k$, this expression is
useful for computational approximation even for small values of $k$.

In this paper, we give exact expressions for the Stieltjes constants as double
infinite series, and one of the series has an exponentially fast rate of convergence.
In a particular case for $a=1$, we are able to recover a result that corresponds
to an earlier series representation for the Riemann zeta function given by Brun
\cite{brun}.  Brun performed `horizontal' and `vertical' summations, and so our
approach is very different.

{\bf Proposition 1}.  Let Re $a>0$.  Then we have 
$$\gamma_0(a)=-\psi(a)=-\ln a+{1 \over a}+\sum_{n=1}^\infty \sum_{j=1}^\infty
{{(-1)^j} \over {j+a2^n}}, \eqno(1.3)$$
and for $\ell \geq 1$
$$\gamma_\ell(a)=-{{\ln^{\ell+1} a} \over {\ell+1}}+{1 \over a}\ln^\ell a+ 
\sum_{n=1}^\infty \sum_{j=1}^\infty {{(-1)^j} \over {j+a2^n}}\ln^\ell\left(a+{j \over
2^n}\right). \eqno(1.4)$$

As a byproduct of our proof of Proposition 1, we obtain the following double 
series representation.  We have
{\newline \bf Corollary 1}.  For Re $s>1$ and Re $a>0$, we have
$$\zeta(s,a)=a^{-s}+{a^{1-s} \over {s-1}}+\sum_{n=1}^\infty 2^{n(s-1)} \sum_{j=1}^\infty
{{(-1)^j} \over {(j+a2^n)^s}}.  \eqno(1.5)$$

Brun \cite{brun} realized the following representation of the Riemann zeta function for
Re $s>1$:
$$\zeta(s)={1 \over {s-1}}+1-\beta(s), \eqno(1.6a)$$
where 
$$\beta(s)=\sum_{n,j=1}^\infty {{(-1)^{j-1}2^{n(s-1)}} \over {(2^n+j)^s}}.  \eqno(1.6b)$$
Therefore, we recover his result as the special case at $a=1$.

{\bf Corollary 2}.  We have
$$\gamma={1 \over 2}\sum_{n=0}^\infty \left[\psi\left({{2^{n+1}+1} \over 2}\right)
-\psi(2^n)\right]$$
$$=\sum_{n=0}^\infty [\psi(2^{n+1})-\psi(2^n)-\ln 2].  \eqno(1.7)$$

{\bf Corollary 3}.  We have for Re $a>0$ 
$$\gamma_0(a)=-\ln a+\sum_{n=0}^\infty[\psi(a2^{n+1})-\psi(a2^n)-\ln 2].  \eqno(1.8)$$
This results agrees with (2.24) of \cite{coffeyjnt}.


{\bf Proposition 2}.  
We have for $\ell \geq 0$, Re $a>0$, and integers $k \geq 2$
$$\gamma_\ell(a)={1 \over 2}{{\ln^\ell a} \over a}-{{\ln^{\ell+1}a} \over {\ell+1}}-\sum_{n=0}^\infty {1 \over k^n}\sum_{j=0}^\infty \left\{{1 \over 2} \left({1 \over k}-1\right)\left[{{\ln^\ell(bj+a)} \over {(bj+a)}} + {{\ln^\ell(b(j+1)+a)} \over {(b(j+1)+a)}}\right]\right.$$
$$\left.+{1 \over k}\sum_{m=1}^{k-1} {{\ln^\ell[b(j+m/k)+a]} \over {[b(j+m/k)+a]}}\right\}_{b=k^{-n}}  \eqno(1.9a)$$
$$=-{{\ln^{\ell+1}a} \over {\ell+1}}+{1 \over k}\sum_{n=0}^\infty\sum_{m=1}^{k-1} \left. {{\ln^\ell[b m/k+a]} \over {m/k+ak^n}}\right|_{b=k^{-n}}$$
$$-\sum_{n=0}^\infty \sum_{j=1}^\infty \left\{
\left({1 \over k}-1\right){{\ln^\ell(bj+a)} \over {j+ak^n}}
+{1 \over k}\sum_{m=1}^{k-1} {{\ln^\ell[b(j+m/k)+a]} \over {j+m/k+ak^n}}\right\}_{b=k^{-n}}.  \eqno(1.9b)$$

Our results have application to Dirichlet $L$ functions, as these may be written as a linear combination of Hurwitz zeta functions.  For instance, for $\chi$ a principal (nonprincipal) character modulo $m$ and Re $s > 1$ (Re $s > 0$) we have
$$L(s,\chi) = \sum_{k=1}^\infty {{\chi(k)} \over k^s} ={1 \over m^s}\sum_{k=1}^m 
\chi(k) \zeta\left(s,{k \over m}\right).  \eqno(1.10)$$

\pagebreak
\centerline{\bf Proof of Propositions}

{\it Proposition 1}.  We apply
{\newline \bf Lemma 1}.  We have for Re $s>1$,
$$\zeta(s,a)-{a^{1-s} \over {s-1}}={a^{-s} \over 2}+{1 \over 4}\sum_{n=0}^\infty 2^{-n}
\sum_{j=0}^\infty \left[{1 \over {(2^{-n}j+a)^s}}+ {1 \over {(2^{-n}j+2^{-n}+a)^s}}-
{2 \over {(2^{-n}(j+1/2)+a)^s}}\right]$$
$$=a^{-s}+{1 \over 2}\sum_{n=0}^\infty 2^{n(s-1)}\sum_{j=0}^\infty \left[{1 \over {(j+
1+a2^n)^s}}-{1 \over {(j+1/2+a2^n)^s}}\right].  \eqno(2.1)$$

We start from the representation
$$\zeta(s,a)={a^{-s} \over 2}+{a^{1-s} \over {s-1}}-s\int_0^\infty {{P_1(x)} \over
{(x+a)^{s+1}}} dx, ~~~~\mbox{Re} ~s >0, \eqno(2.2)$$
and employ the functions \cite{coffeyjnt} $f(x)=-P_1(x)$ and $g_2(x)=f(x)-f(2x)/2$,
with $\sum_{n=0}^\infty g_2(2^n x)/2^n=f(x)$.  We find
$$\zeta(s,a)={a^{-s} \over 2}+{a^{1-s} \over {s-1}}+s\sum_{n=0}^\infty 2^{-n} \int_0^\infty {{g_2(2^nx)} \over {(x+a)^{s+1}}} dx$$
$$={a^{-s} \over 2}+{a^{1-s} \over {s-1}}+s\sum_{n=0}^\infty 4^{-n} \sum_{j=0}^\infty
\int_j^{j+1} {{g_2(y)dy} \over {(2^{-n}y+a)^{s+1}}}$$
$$={a^{-s} \over 2}+{a^{1-s} \over {s-1}}+{s\over 4}\sum_{n=0}^\infty 4^{-n} \sum_{j=0}^\infty \left(\int_j^{j+1/2}-\int_{j+1/2}^{j+1}\right) {{dy} \over {(2^{-n}y+a)^{s+1}}}.  \eqno(2.3)$$
Carrying out the integrations then gives the first line of (2.1).  Elementary
manipulations then yield
$$\zeta(s,a)={a^{-s} \over 2}+{a^{1-s} \over {s-1}}+{1 \over 4}\sum_{n=0}^\infty 2^{n(s-1)}\sum_{j=0}^\infty \left[{1 \over {(j+1+a2^n)^s}}+{1 \over {(j+a2^n)^s}}
-{2 \over {(j+1/2+a2^n)^s}}\right]$$
$$={a^{-s} \over 2}+{a^{1-s} \over {s-1}}+{1 \over 4}\sum_{n=0}^\infty 2^{n(s-1)}\left\{{1 \over {a^s2^{ns}}}+2\sum_{j=0}^\infty \left[{1 \over {(j+1+a2^n)^s}}
-{1 \over {(j+1/2+a2^n)^s}}\right]\right\},  \eqno(2.4)$$
giving the rest of the Lemma.  We remark that (2.3) is generalized in (2.14) for 
values of $k \geq 2$.

We now use standard expansions about $s=1$, including
$$(j+\beta+a2^n)^{s-1+1}=(j+\beta+a2^n)\exp[(s-1)\ln(j+\beta+a2^n)], \eqno(2.5)$$
to write
$$\zeta(s,a)-{1 \over {s-1}}=\sum_{\ell=1}^\infty {{(-1)^\ell} \over {\ell!}}\ln^\ell a
(s-1)^{\ell-1}+{1 \over a}\sum_{\ell=0}^\infty {{(-1)^\ell} \over {\ell!}}\ln^\ell a
(s-1)^{\ell}$$
$$+{1 \over 2}\sum_{n=0}^\infty \sum_{j=0}^\infty \sum_{\ell=0}^\infty {{(s-1)^\ell} 
\over {\ell !}}\left\{{{[n\ln 2-\ln(j+1+a2^n)]^\ell} \over {j+1+a2^n}}-
{{[n\ln 2-\ln(j+1/2+a2^n)]^\ell} \over {j+1/2+a2^n}}\right\}.  \eqno(2.6)$$
Comparing to the expansion (1.1) we have
$$\gamma_\ell(a)=-{{\ln^{\ell+1} a} \over {\ell+1}}+{1 \over a}\ln^\ell a$$
$$+{{(-1)^\ell} \over 2}\sum_{n=0}^\infty \sum_{j=0}^\infty \left\{{{[n\ln 2-\ln(j+1+a2^n)]^\ell} \over {j+1+a2^n}}-
{{[n\ln 2-\ln(j+1/2+a2^n)]^\ell} \over {j+1/2+a2^n}}\right\}$$  
$$=-{{\ln^{\ell+1} a} \over {\ell+1}}+{1 \over a}\ln^\ell a
+\sum_{n=0}^\infty \sum_{j=0}^\infty \left[{{\ln^\ell\left(a+{{j+1} \over 2^n}\right)
} \over {2j+2+a2^{n+1}}} -{{\ln^\ell\left(a+{{j+1/2} \over 2^n}\right)
} \over {2j+1+a2^{n+1}}}\right]$$
$$=-{{\ln^{\ell+1} a} \over {\ell+1}}+{1 \over a}\ln^\ell a
+\sum_{n=1}^\infty \sum_{j=1}^\infty \left[{{\ln^\ell\left(a+{j \over 2^{n-1}}\right)
} \over {2j+a2^n}} -{{\ln^\ell\left(a+{{j-1/2} \over 2^{n-1}}\right)} \over {2j-1+a2^n}}\right].  \eqno(2.7)$$
This completes the Proposition.

We can rearrange (2.1) in the form
$$\zeta(s,a)-{a^{1-s} \over {s-1}}=a^{-s}+{1 \over 2}\sum_{n=1}^\infty 2^{(n-1)(s-1)}
2^s \sum_{j=1}^\infty \left[{1 \over {(2j+a^n)^s}}-{1 \over {(2j-1+a^n)^s}}\right].
\eqno(2.8)$$
From this follows Corollary 1.

{\it Remark}.  It is easily verified that from Corollary 1 we may obtain the property
$\partial_a\zeta(s,a)=-s\zeta(s+1,a)$.  Likewise it follows that $\zeta(s,1/2)=(2^s-1)\zeta(s)$, and we provide a demonstration of this property 
from (1.5).  We have
$$\zeta\left(s,{1 \over 2}\right)=2^s+{2^{s-1} \over {s-1}}+\sum_{n=1}^\infty
2^{n(s-1)} \sum_{j=1}^\infty {{(-1)^j} \over {(j+2^{n-1})^s}}.  \eqno(2.9)$$
We transform the latter double sum to
$$\sum_{n=0}^\infty 2^{(n+1)(s-1)} \sum_{j=1}^\infty {{(-1)^j} \over {(j+2^n)^s}}
=2^{s-1}\left[\sum_{j=1}^\infty {{(-1)^j} \over {(j+1)^s}}+\sum_{n=1}^\infty 2^{n(s-1)} \sum_{j=1}^\infty {{(-1)^j} \over {(j+2^n)^s}}\right]$$
$$=2^{s-1}\left[\zeta(s)-1-2^{1-s}\zeta(s)+\sum_{n=1}^\infty 2^{n(s-1)} \sum_{j=1}^\infty {{(-1)^j} \over {(j+2^n)^s}}\right].  \eqno(2.10)$$
Then from (2.9) and (1.5) with $a=1$,
$$\zeta\left(s,{1 \over 2}\right)=2^s+{2^{s-1} \over {s-1}}+2^{s-1}\zeta(s)-2^{s-1}
-\zeta(s)
+2^{s-1}\sum_{n=1}^\infty 2^{n(s-1)} \sum_{j=1}^\infty {{(-1)^j} \over {(j+2^n)^s}}$$
$$=2^s+{{2^{s-1}-2^{s-1}} \over {s-1}}+2^s\zeta(s)-2^s-\zeta(s)$$
$$=(2^s-1)\zeta(s).  \eqno(2.11)$$
Above, we used the alternating form of the zeta function so that
$$\sum_{j=1}^\infty {{(-1)^j} \over {(j+1)^s}}=-\left(\sum_{j=1}^\infty {{(-1)^j} 
\over j^s} + 1\right)=(1-2^{1-s})\zeta(s)-1.  \eqno(2.12)$$

{\it Corollary 2}.  The first expression for $\gamma$ follows by performing one of the sums in the Brun result (1.6), 
$$\gamma=1-\beta(1)=1-\sum_{n=1}^\infty \sum_{j=1}^\infty {{(-1)^{j-1}} \over 
{2^n+j}}.  \eqno(2.13)$$
In order to obtain the second expression for $\gamma$, we apply the duplication
formula of the digamma function, $2\psi(2x)=2\ln 2+\psi(x)+\psi(x+1/2)$.

{\it Remark}.  The second expression for $\gamma$ in Corollary 2 is precisely
(2.24) of \cite{coffeyjnt} with $a=1$.

{\it Corollary 3}.  We have from (1.3)
$$\gamma_0(a)=-\psi(a)=-\ln a +{1 \over a}+{1 \over 2}\sum_{n,j=0}^\infty \left[{1 
\over {j+1+a2^n}}-{1 \over {j+1/2+a2^n}}\right].  \eqno(2.14)$$
Summing over $j$ then gives
$$\gamma_0(a)=-\ln a +{1 \over a}+{1 \over 2}\sum_{n=0}^\infty \left[\psi\left(a2^n+
{1 \over 2}\right)-\psi\left(a2^n+1\right)\right].  \eqno(2.15)$$
We then use both the functional equation $\psi(x+1)=\psi(x)+1/x$ and the 
duplication formula of the digamma function to obtain
$$\gamma_0(a)=-\ln a +{1 \over a}+\sum_{n=0}^\infty \left[\psi\left(a2^{n+1}
\right)-\psi\left(a2^n\right)-\ln 2-{1 \over {a2^{n+1}}}\right].  \eqno(2.16)$$
The Corollary follows.

{\it Proposition 2}.  We simply outline the proof.  We start again from the
integral representation (2.2).  Now we make use of the functions for $k \geq 2$ and $f(x)=-P_1(x)$ \cite{coffeyjnt},
$$g_k(x)=f(x)-{1 \over k}f(kx),  \eqno(2.17)$$
with $\sum_{n=0}^\infty {{g_k(k^n x)} \over k^n}=f(x)$.  
Then
$$\zeta(s,a)-{a^{-s} \over 2}-{a^{1-s} \over {s-1}}$$
$$=s\sum_{n=0}^\infty {1 \over k^n}\int_0^\infty {{g_k(k^nx)} \over {(x+a)^{s+1}
}}dx$$
$$=s\sum_{n=0}^\infty {1 \over k^{2n}}\int_0^\infty {{g_k(y)dy} \over {(k^{-n}y+a)^{s+1} }}$$
$$=s\sum_{n=0}^\infty {1 \over k^{2n}}\sum_{j=0}^\infty \int_j^{j+1} {{g_k(y)dy} \over {(k^{-n}y+a)^{s+1} }}$$
$$=s\sum_{n=0}^\infty {1 \over k^{2n}}\sum_{j=0}^\infty \left[{1 \over 2}\left(1-
{1 \over k}\right)\int_j^{j+1/k} + {1 \over 2}\left(1-{3 \over k}\right) \int_{j+1/k}^{j+2/k} + {1 \over 2}\left(1- {5 \over k}\right)\int_{j+2/k}^{j+3/k} 
\right.$$
$$\left. + \ldots + {1 \over 2}\left({1 \over k}-1\right)\int_{j+(k-1)/k}^{j+1} 
\right]{{dy} \over {(k^{-n}y+a)^{s+1} }}.  \eqno(2.18)$$
In the last step we have used the values of $g_k(x)$ on subintervals $\left[{{j-1} \over k},{j \over k} \right)$ for $j=1,2,\ldots,k$, 
$$g_k(x)={1 \over 2}\left(1-{1 \over k}\right)-{{(j-1)} \over k}, ~~~~~~x \in \left[{{j-1} \over k},{j \over k}\right).  \eqno(2.19)$$
In particular, the difference of these values on consecutive subintervals is
simply $1/k$.  We then perform the integrations and collect the terms to find for Re $s>1$
$$\zeta(s,a)={a^{-s} \over 2}+{a^{1-s} \over {s-1}}$$
$$=-\sum_{n=0}^\infty {1 \over k^n}\sum_{j=0}^\infty \left\{{1 \over 2}\left({1 \over k}-1\right)\left[{1 \over {(bj+a)^s}} + {1 \over {(b(j+1)+a)^s}}\right]+{1 \over k}\sum_{m=1}^{k-1} {1 \over {[b(j+m/k)+a]^s}}\right\}_{b=k^{-n}}.  \eqno(2.20)$$ 
Expanding about $s=1$ and using the definition (1.1) gives the Proposition.

\bigskip
\centerline{\bf Summary}
\medskip

Among other results, we have found computationally useful series representations
of the Stieltjes constants.  Although the series representations are doubly infinite,
one of the series converges exponentially quickly with parameter $k \geq 2$.  We 
have obtained in (2.16) a parameterized representation of the Hurwitz zeta function.
In the very special case of $k=2$ and $a=1$, we recover the much earlier result 
of Brun \cite{brun} of a double series expression for the Riemann zeta function.
Since $\gamma_0(a)=-\psi(a)$, where $\psi$ is the digamma function, even at the
lowest order, we effectively have found series representations of the harmonic
numbers, generalized harmonic numbers, and other mathematical constants.  We recall
that the generalized harmonic numbers $H_n^{(r)} \equiv \sum_{k=1}^n 1/k^r$ may be
readily found from the polygamma functions $\psi^{(r-1)}$.  Very special limit cases of
our results include parameterized series representations for the expressions
$\zeta(0,a)=1/2-a$ and $\zeta'(0,a)=\ln \Gamma(a)-(1/2)\ln(2\pi)$.  Further, our
results have application to series representation and expansions of Dirichlet $L$
functions.



\pagebreak


\begin{thebibliography}{99}
\bibitem{nbs}M. Abramowitz and I. A. Stegun,
{Handbook of Mathematical Functions, Washington, National Bureau of Standards
(1964).}
\bibitem{addison}A. W. Addison,
{A series representation for Euler's constant, Amer. Math. Monthly {\bf 74}, 
823-824 (1967).}
\bibitem{andrews}G. E. Andrews, R. Askey, and R. Roy, 
{Special Functions, Cambridge University Press (1999).}
\bibitem{brun}V. Brun,
{Deux transformations \'{e}l\'{e}mentaires de la fonction Zeta de Riemann, Rivista Ci. Lima {\bf 41}, 517-525 (1939).}
\bibitem{coffeyjnt}M. W. Coffey,
{Addison-type series representation for the Stieltjes constants, J. Number Th. {\bf 130}, 2049-2064 (2010); arXiv:0912.2391 (2009).}
\bibitem{coffeyjmaa}M. W. Coffey,
{New results on the Stieltjes constants:  Asymptotic and exact evaluation, 
J. Math. Anal. Appl. {\bf 317}, 603-612 (2006); arXiv:math-ph/0506061.}
\bibitem{coffeystdiffs}M. W. Coffey,
{On representations and differences of Stieltjes coefficients, and other relations, to appear in Rocky Mtn. J. Math.; arXiv/math-ph/0809.3277v2 (2008).}
\bibitem{coffey2009}M. W. Coffey,
{Series representations for the Stieltjes constants, arXiv:0905.1111 (2009).}
\bibitem{edwards}H. M. Edwards,
{Riemann's Zeta Function, Academic Press, New York (1974).}
\bibitem{grad}I. S. Gradshteyn and I. M. Ryzhik,
{Table of Integrals, Series, and Products, Academic Press, New York (1980).}
\bibitem{ivic}A. Ivi\'{c}, 
{The Riemann Zeta-Function, Wiley, New York (1985).}
\bibitem{knessl2}C. Knessl and M. W. Coffey,
{An asymptotic form for the Stieltjes constants $\gamma_k(a)$ and for a sum $S_\gamma(n)$ appearing under the Li criterion (2010), to appear in Math. Comp.}
\bibitem{riemann}B. Riemann,
{\"{U}ber die Anzahl der Primzahlen unter einer gegebenen Gr\"{o}sse, 
Monats. Preuss. Akad. Wiss., 671 (1859-1860).}
\bibitem{stieltjes}T. J. Stieltjes,
{Correspondance d'Hermite et de Stieltjes, Volumes 1 and 2, Gauthier-Villars,
Paris (1905).}
\bibitem{titch}E. C. Titchmarsh,
{The Theory of the Riemann Zeta-Function, 2nd ed., Oxford University
Press, Oxford (1986).}
\bibitem{wilton2}J. R. Wilton, 
{A note on the coefficients in the expansion of $\zeta(s,x)$ in powers of 
$s-1$, Quart. J. Pure Appl. Math. {\bf 50}, 329-332 (1927).}
\end{thebibliography}
\end{document}